# Experimental Realization of Geometry-Dependent Skin Effect in a Reciprocal Two-Dimensional Lattice

Wei Wang[*,2], Mengying Hu[*,1], Xulong Wang[2], Guancong Ma[2], Kun Ding[1]

[1]Department of Physics, State Key Laboratory of Surface Physics, and Key Laboratory of Micro and Nano Photonic Structures (Ministry of Education), Fudan University, Shanghai 200438, China

[2]Department of Physics, Hong Kong Baptist University, Kowloon Tong, Hong Kong, China

*These authors contributed equally.

**Abstract**

Recent studies of non-Hermitian periodic lattices unveiled the non-Hermitian skin effect (NHSE), in which the bulk modes under the periodic boundary conditions (PBC) become skin modes under open boundary conditions (OBC). The NHSE is a topological effect owing to the non-trivial spectral winding, and such spectral behaviors appear naturally in non-reciprocal systems. Hence prevailing approaches rely on non-reciprocity to achieve the NHSE. Here, we report the experimental realization of the geometry-dependent skin effect (GDSE) in a two-dimensional (2D) reciprocal system, in which the skin effect occurs only at boundaries whose macroscopic symmetry mismatches with the lattice symmetry. The role of spectral reciprocity and symmetry is revealed by connecting reflective channels at given boundaries with the spectral topology of the PBC spectrum. Our work highlights the vital role of reciprocity and macroscopic geometry on the NHSE in systems with dimensions larger than one and opens new routes for wave structuring using non-Hermitian effects.

***Introduction***.— The non-Hermitian degrees of freedom [1,2] have significantly broadened the context of Hamiltonian because, unlike Hermitian Hamiltonians that exclusively treat closed systems with steady states, non-Hermitian effective Hamiltonians can handle a large variety of open systems [3–5]. Such a generalization necessitates re-examining the basic concepts and relations built on the Hermitian formalism, such as the bulk-edge correspondence [6,7] and spectral degeneracy [8,9]. A critical difference is that the non-Hermitian spectrum lies in the complex plane, giving rise to spectral loops and point gaps [10–13], as well as defective spectral degeneracies [14–20]. The research into these exclusive characteristics has led to the discovery of a plethora of unique phenomena [5,21–23]. Perhaps one of the most prominent phenomena is the non-Hermitian skin effect (NHSE), which causes the bulk modes to collapse towards the open boundary of the system [24–30,46]. NHSE was discovered by studying one-dimensional (1D) non-reciprocal lattices, wherein the Bloch Hamiltonian has a non-reciprocal spectrum, i.e., $\omega(\boldsymbol{k}) \neq \omega(-\boldsymbol{k})$ [25,26,31]. It has been confirmed by various experiments in 1D and 2D systems [32–37].

Recent studies further show that NHSE can be enriched by the consideration of symmetry. For example, anomalous time-reversal symmetry (TRS$^\dagger$) can lead to the $\mathbb{Z}_2$ skin effect in one-dimensional lattices [29,38], for which the spectrum under the periodic boundary condition (PBC) is reciprocal, i.e., $\omega(k) = \omega(-k)$. This new classification also implies that non-reciprocal or directional hopping is not always necessary for the NHSE. Extending to higher-dimensional systems, it is claimed that the NHSE can universally appear in both non-reciprocal and reciprocal systems [39,40]. Unlike the NHSE in non-reciprocal systems, the combination of the lattice symmetry and spectral reciprocity can lead to the disappearance of skin modes at some particular boundaries, so the skin effect is said to be dependent on the geometry of the open-boundary lattice [39,41]. However, the experimental confirmation of this GDSE so far remains absent.

In this work, we study a 2D reciprocal model wherein the only form of non-Hermiticity is the dissipative rate. By establishing a connection between the geometry of the open boundary and the winding behavior of the PBC spectrum, we identify that the skin effect in such a system only occurs at boundaries that are not parallel to the primitive vectors of the lattice. Such skin effects and the geometry-dependent characteristic are experimentally realized using active mechanical lattices.

***Theoretical model***.— To elucidate the 2D GDSE, we first recall the physics of the non-reciprocal NHSE in 1D open lattices. The PBC spectrum, denoted $\omega^P(k)$, is a complex function that maps the BZ to a closed loop in the 2D complex energy manifold. When such spectral loops enclose a nonzero area, the wavevectors $k$ and $-k$ are mapped to different complex energy. This has profound consequences when OBC is considered: when a wave with a wavevector $k$ impinges on an open boundary, there is no reflective channel with the wavevector $-k$ at the same energy. As a result, the energy carried by the incident wave has no choice other than accumulating at the boundary, leading to the NHSE. On the other hand, if the area enclosed by the spectrum is zero, i.e., the spectral loop collapses to an arc, implying at any complex energy, there exist two choices of wavevectors, and the NHSE does not emerge. Hence, the winding behavior of $\omega^P(k)$ is a convenient tool for predicting the NHSE: a spectral loop (arc) in $\omega^P(k)$, which has a nonzero (zero) winding number, directly corresponds to the occurrence (absence) of the NHSE.

We next consider 2D non-Hermitian lattices. Intuitively, one would expect that the skin effect should meet a more stringent set of conditions because there is simply "more room" for the waves to move around, or more channels to leave a boundary. However, this is not the case, as NHSE is ubiquitous and emerges even in the absence of non-reciprocity [39]. To understand this, we consider a stacked non-Hermitian Su-Schrieffer-Heeger model shown in Fig. 1(a). The momentum-space Hamiltonian under PBC reads

$$H(k_x, k_y) = 2\pi \begin{pmatrix} f_A - i\gamma_A + 2v_x \cos k_x & v_y + w_y e^{-ik_y} \\ v_y + w_y e^{ik_y} & f_B + i\gamma_B + 2v_x \cos k_x \end{pmatrix}, \quad (1)$$

wherein all parameters in Eq. (1) are real. $v_y$ ($w_y$) is the intracell (intercell) hopping in the $y$-direction, and $v_x$ is the hopping along the $x$-direction. The non-Hermiticity is due to $-i\gamma_A$ and $i\gamma_B$, which are the typical onsite loss and gain. Model (1) respects reciprocity because it clearly satisfies $H(k_x, k_y) = H^T(-k_x, -k_y)$. The PBC spectrum is obviously reciprocal, i.e., $\omega^P(k_x, k_y) = \omega^P(-k_x, -k_y)$ [Figs. 1(b) and (c)]. In addition, it is easy to check that the spectrum is symmetric in both $k_x$ and $k_y$, i.e., $\omega^P(k_x, k_y) = \omega^P(-k_x, k_y) = \omega^P(k_x, -k_y)$, as shown in Figs. 1(b) and (c). Consider the case shown in Fig. 1(d): an incident wave with a wavevector $\boldsymbol{q}_{in}(\omega) = (q_\parallel^{in}, q_\perp^{in})^T$ encounters an open boundary, where $q_\parallel^{in}$ and $q_\perp^{in}$ are the wavevector components parallel and perpendicular to the boundary, respectively. Because the parallel wavevector is conserved for both the incident and reflected waves, i.e., $q_\parallel^{in} = q_\parallel^{re} = q_\parallel$, the skin effect would occur when the reflected wave has no perpendicular component $q_\perp^{re}$ at the same $\omega$. (We do not consider situations that $q_\parallel$ is different for the incoming and outgoing waves, such as metamaterials with gradient indices at the subwavelength scale.) In other words, the condition of NHSE is the existence of two different $q_\perp$ as the pre-images for a particular $\omega$.

In the general case, the real-space boundary is not parallel with either of the primitive vectors. Such a boundary determines the local coordinate of $(\hat{q}_\parallel, \hat{q}_\perp)$ in the momentum space, which is linked to $(\hat{k}_x, \hat{k}_y)$ by a frame rotation [Figs. 1(d, e)]. In other words, the boundary fixes the slope of a series of lines crossing the Brillouin zone (BZ), such as the red dashed lines in Fig. 1(e). Each line in the $\hat{q}_\perp$-direction is a closed parametric loop in the momentum space, and they are mapped to the PBC spectrum as loops by the function $\omega^P(q_\parallel, q_\perp)$. The behavior of these spectral loops can be quantified by a spectral winding number

$$\mathcal{W}_{q_\parallel} \equiv \frac{1}{2\pi i} \int_{\Gamma_{q_\parallel}} dq_\perp (\partial_{q_\perp} \ln \det[H(q_\parallel, q_\perp) - \omega_b]), \qquad (2)$$

where $\omega_b \in \mathbb{C}$ is a reference spectral point. Note that Eq. (2) integrates over $q_\perp$ along $\Gamma_{q_\parallel}$, which is a closed loop [the solid blue line in Fig. 1(e), which is equivalent to the red dashed lines]. For any $q_\parallel$, if $\mathcal{W}_{q_\parallel} \neq 0$, which indicates that the spectral loop encloses a nonzero spectral area, no two solutions of $q_\perp$ can be found at the same energy. Skin effects must occur. In contrast, if $\forall q_\parallel$ in BZ, $\mathcal{W}_{q_\parallel} = 0$, i.e., the spectral loops have no interior, so at each energy, the incident and reflected waves can have different $q_\perp$. As a result, the reflected channel exists at the boundary, and the skin effect is absent. In Fig. 1(f), we show the PBC spectrum as a function of $q_\parallel$ for the oblique edge. Spectral loops enclosing nonzero areas clearly exist for most $q_\parallel$, e.g., at $q_\parallel = 0.5\pi$ and $q_\parallel = -0.6\pi$. Hence the skin effect occurs at this edge. However, if the boundary cuts parallelly to the $y$-direction, as shown in Fig. 1(h-i), the spectral loops collapse to arcs and enclose zero areas for all $q_\parallel$, so $\mathcal{W}_{q_\parallel}$ always vanishes [Fig. 1(j)]. Thus, the skin effect must be absent at this boundary. This can also be seen by the fact that the PBC spectrum obeys $\omega^P(k_x, k_y) = \omega^P(-k_x, k_y)$, so for any given $k_y$ there always exists a pair of wavevectors $k_x$ and $-k_x$, which means the reflective channel always exists. Since $\omega^P(k_x, k_y) = \omega^P(k_x, -k_y)$, the same can be deduced for the situation that the boundary is parallel to the $x$-direction. In summary, the presence or absence of the skin effect depends on the boundary orientation of the OBC lattice or, equivalently, the geometry of the OBC lattice. This is the GDSE. The GDSE is also characterized by the vanishing of a current functional in the PBC spectra, which makes it fundamentally different from the non-reciprocal NHSE [42]. The GDSE is also different from the NHSE induced by non-reciprocity by having different volume laws, i.e., the relation between the number of skin modes and the lattice size [42].

Interestingly, owing to the spectral reciprocity, when $q_\parallel = 0$, the condition $\omega^P(0, q_\perp) = \omega^P(0, -q_\perp)$ is always met. It follows that $\mathcal{W}_{q_\parallel=0} = 0$, as shown by the blue curve in Fig. 1(f). This implies that reflective channels always exist for the normal-incident waves, regardless of the orientation of the boundary $\hat{q}_\parallel$. As a result, even when the GDSE occurs, a portion of modes must remain extended bulk modes, which would be clearly identified in the thermodynamic limit. In contrast, if spectral reciprocity breaks, then $\forall q_\parallel$ in the BZ, $\mathcal{W}_{q_\parallel} \neq 0$, so all modes are skin modes. (Exceptions such as the modes at the "Bloch points" may exist, but those are not generic cases and usually require parameter tuning.)

***Experimental realization.***— Equation (1) can be realized in an active mechanical lattice [37,43]. The unit cell consisting of two rotational oscillators coupled by tension springs is shown in Fig. 2(a). The non-Hermitian terms, i.e., onsite loss $(-i\gamma_A)$ and gain $(i\gamma_B)$, are achieved by imposing a self-feedback torque $\tau(t) = \beta\dot{\theta}(t)$ on every oscillator [42]. Here, $\dot{\theta}(t)$ is the instantaneous angular velocity and $\beta$ is a tunable constant. When $\beta$ is positive (negative), $\tau(t)$ acts as effective gain (loss). However, we found that the active torque also introduces a $\beta$-dependent perturbation to the onsite resonance, as shown in Fig. S4(b). This is accounted by $f_A$ and $f_B$ in Eq. (1). All coefficients in Eq. (1) are obtained by Green's function from the measured data [16] (see Ref. [42] for the retrieved values).

Our experimental system is a trapezoid lattice consisting of $N = 78$ sites, as shown in Figs. 2(b) and (d). And the corresponding OBC spectrum, denoted $\omega^O$, is shown in Fig. 2(c). According to the above analysis, the skin effect should occur only at the oblique edge. This is confirmed in Fig. 2(b), wherein the colormap depicts the spatial distribution of all eigenstates, defined as $\Psi(\mathbf{x}) = \frac{1}{N}\sum_n |\psi_n(\mathbf{x})|^2$, where $\psi_n(\mathbf{x})$ is the $n$-th normalized right eigenstate. To quantify the degree of localization of each mode, we introduce the quantity

$$\Upsilon_n = \frac{\sum_{\mathbf{x}\in\mathcal{B}} |\psi_n(\mathbf{x})|^2}{\sum_{\mathbf{x}} |\psi_n(\mathbf{x})|^2}, \tag{3}$$

where $\mathcal{B}$ denotes the oblique boundary region with the width of a single unit cell [indicated by the gray dashed line in Fig. 2(b)]. A large $\Upsilon_n$ indicates the corresponding wavefunction is strongly localized in $\mathcal{B}$. In Fig. 2(c), the color of the dots represents $\Upsilon_n$. Strongly localized modes are found near $\text{Re}(\omega^0/2\pi) \in (12.0, 13.1)$ Hz. These modes lie near $\text{Im}(\omega^0) \cong 0$, so that they can be stably excited with minimal temporal decay. Experimentally, a source placed at the right edge sends a pulsing containing $\Delta f_1 = [12.0, 13.1]$ Hz. The spectral responses $A_u(f)$ are measured across the entire lattice, where $u$ denotes the site number. We then compute the spectral sum

$$P_u = \int_{\Delta f_1} |A_u(f)|^2 df. \tag{4}$$

The result shows a drop in $P_u$ immediately away from the source, but then it increases again and peaks at the oblique edge [Fig. 2(f)], which is strong evidence of the skin effect. No localization is observed at the boundaries parallel to the $x$ and $y$ directions.

In a control experiment, we excite the lattice at the same position but with a pulse containing $\Delta f_2 = [9.0, 11.0]$ Hz. The modes in this frequency interval have small $\Upsilon_n$, but their imaginary eigenfrequencies are negative $\text{Im}(\omega^0) < 0$, which indicates a temporal decaying characteristic. Their responses are shown in Fig. 2(e), which indeed decay rapidly away from the source. These results conform well with the predictions of the GDSE. (The modes lying within $\text{Re}(\omega^0/2\pi) \in (11.0, 12.0) \cup (13.1, 14.1)$ Hz have $\text{Im}(\omega^0) > 0$. They are temporally unstable and hence are not excited by purposely avoiding the corresponding frequencies.)

For further validation, we perform similar experiments on the same system arranged in a rectangular lattice consisting of 70 sites [Figs. 3(a) and (c)]. In Fig. 3(a), $\Psi(\mathbf{x})$ is clearly uniform, which indicates the absence of the skin effect. Figure 3(b) plots the PBC and OBC spectra. The color of the OBC spectrum represents $\Upsilon_n$ with the region $\mathcal{B}$ now being the left edge. Compared with Fig. 3(c), all $\Upsilon_n$ are close to zero, which suggests that the modes are extended. In the experiment, the source is placed in the bulk, and the excitation covers 9.5–15.0 Hz. The spectral sum plotted in Fig. 3(d) shows the typical behavior of the bulk modes under dissipation. No localization at the other boundaries occurs.

***Conclusion.*** — In summary, we have demonstrated the GDSE in a 2D reciprocal mechanical system in which the only non-Hermiticity is the gain and loss. The origin of the GDSE is an interplay of the PBC spectrum, lattice symmetry, and macroscopic symmetry of the lattice. In other words, the skin effect occurs in generic geometric shapes as long as the macroscopic geometry does not align with any lattice symmetries. Equivalently, the persistence of extended bulk modes against the non-Hermiticity requires the combined protection of the reciprocity and the lattice symmetries, hence it is not a generic case. Therefore, our results not only reveal the universality of the skin effect in systems with dimensions higher than one but also highlight the vital role of macroscopic geometry and symmetry for the skin effects. Recent works have also predicted that the interplay between geometry and symmetry can dramatically alter the generalized BZ of 2D systems, including its shapes and even dimensionality [44,45]. Thus, the rich diversity of symmetries available in higher dimensions is a potential arsenal of powerful tools for discovering non-Hermitian effects and tailoring their applications.

*Acknowledgment.*— This work is supported by the National Natural Science Foundation of China (12174072, 11922416), the Hong Kong Research Grants Council (RFS2223-2S01, 12302420, 12300419, 12301822), and the Natural Science Foundation of Shanghai (No. 21ZR1403700).

**References**

[1] C. M. Bender and S. Boettcher, *Real Spectra in Non-Hermitian Hamiltonians Having P T Symmetry*, Phys. Rev. Lett. **80**, 5243 (1998).

[2] C. M. Bender, D. C. Brody, and H. F. Jones, *Complex Extension of Quantum Mechanics*, Phys. Rev. Lett. **89**, 270401 (2002).

[3] C. M. Bender, *Making Sense of Non-Hermitian Hamiltonians*, Rep. Prog. Phys. **70**, 947 (2007).

[4] I. Rotter, *A Non-Hermitian Hamilton Operator and the Physics of Open Quantum Systems*, J. Phys. Math. Theor. **42**, 153001 (2009).

[5] Y. Ashida, Z. Gong, and M. Ueda, *Non-Hermitian Physics*, Adv. Phys. **69**, 249 (2020).

[6] M. Z. Hasan and C. L. Kane, Colloquium *: Topological Insulators*, Rev. Mod. Phys. **82**, 3045 (2010).

[7] E. J. Bergholtz, J. C. Budich, and F. K. Kunst, *Exceptional Topology of Non-Hermitian Systems*, Rev. Mod. Phys. **93**, 015005 (2021).

[8] K. Ding, C. Fang, and G. Ma, *Non-Hermitian Topology and Exceptional-Point Geometries*, Nat. Rev. Phys. **4**, 745 (2022).

[9] N. P. Armitage, E. J. Mele, and A. Vishwanath, *Weyl and Dirac Semimetals in Three-Dimensional Solids*, Rev. Mod. Phys. **90**, 015001 (2018).

[10] Z. Gong, Y. Ashida, K. Kawabata, K. Takasan, S. Higashikawa, and M. Ueda, *Topological Phases of Non-Hermitian Systems*, Phys. Rev. X **8**, 031079 (2018).

[11] H. Shen, B. Zhen, and L. Fu, *Topological Band Theory for Non-Hermitian Hamiltonians*, Phys. Rev. Lett. **120**, 146402 (2018).

[12] K. Kawabata, T. Bessho, and M. Sato, *Classification of Exceptional Points and Non-Hermitian Topological Semimetals*, Phys. Rev. Lett. **123**, 066405 (2019).

[13] K. Kawabata, K. Shiozaki, M. Ueda, and M. Sato, *Symmetry and Topology in Non-Hermitian Physics*, Phys. Rev. X **9**, 041015 (2019).

[14] C. Dembowski, H.-D. Gräf, H. L. Harney, A. Heine, W. D. Heiss, H. Rehfeld, and A. Richter, *Experimental Observation of the Topological Structure of Exceptional Points*, Phys. Rev. Lett. **86**, 787 (2001).

[15] C. E. Rüter, K. G. Makris, R. El-Ganainy, D. N. Christodoulides, M. Segev, and D. Kip, *Observation of Parity–Time Symmetry in Optics*, Nat. Phys. **6**, 192 (2010).

[16] K. Ding, G. Ma, M. Xiao, Z. Q. Zhang, and C. T. Chan, *Emergence, Coalescence, and Topological Properties of Multiple Exceptional Points and Their Experimental Realization*, Phys. Rev. X **6**, 021007 (2016).

[17] B. Zhen, C. W. Hsu, Y. Igarashi, L. Lu, I. Kaminer, A. Pick, S.-L. Chua, J. D. Joannopoulos, and M. Soljačić, *Spawning Rings of Exceptional Points out of Dirac Cones*, Nature **525**, 354 (2015).

[18] A. Cerjan, S. Huang, M. Wang, K. P. Chen, Y. Chong, and M. C. Rechtsman, *Experimental Realization of a Weyl Exceptional Ring*, Nat. Photonics **13**, 623 (2019).

[19] W. Tang, X. Jiang, K. Ding, Y.-X. Xiao, Z.-Q. Zhang, C. T. Chan, and G. Ma, *Exceptional Nexus with a Hybrid Topological Invariant*, Science **370**, 1077 (2020).

[20] J. Liu, Z. Li, Z.-G. Chen, W. Tang, A. Chen, B. Liang, G. Ma, and J.-C. Cheng, *Experimental Realization of Weyl Exceptional Rings in a Synthetic Three-Dimensional Non-Hermitian Phononic Crystal*, Phys. Rev. Lett. **129**, 084301 (2022).

[21] L. Feng, R. El-Ganainy, and L. Ge, *Non-Hermitian Photonics Based on Parity–Time Symmetry*, Nat. Photonics **11**, 752 (2017).

[22] M.-A. Miri and A. Alù, *Exceptional Points in Optics and Photonics*, Science **363**, eaar7709 (2019).

[23] Ş. K. Özdemir, S. Rotter, F. Nori, and L. Yang, *Parity–Time Symmetry and Exceptional Points in Photonics*, Nat. Mater. **18**, 783 (2019).

[24] V. M. Martinez Alvarez, J. E. Barrios Vargas, and L. E. F. Foa Torres, *Non-Hermitian Robust Edge States in One Dimension: Anomalous Localization and Eigenspace Condensation at Exceptional Points*, Phys. Rev. B **97**, 121401 (2018).

[25] F. K. Kunst, E. Edvardsson, J. C. Budich, and E. J. Bergholtz, *Biorthogonal Bulk-Boundary Correspondence in Non-Hermitian Systems*, Phys. Rev. Lett. **121**, 026808 (2018).

[26] S. Yao and Z. Wang, *Edge States and Topological Invariants of Non-Hermitian Systems*, Phys. Rev. Lett. **121**, 086803 (2018).

[27] D. S. Borgnia, A. J. Kruchkov, and R.-J. Slager, *Non-Hermitian Boundary Modes and Topology*, Phys. Rev. Lett. **124**, 056802 (2020).

[28] K. Zhang, Z. Yang, and C. Fang, *Correspondence between Winding Numbers and Skin Modes in Non-Hermitian Systems*, Phys. Rev. Lett. **125**, 126402 (2020).

[29] N. Okuma, K. Kawabata, K. Shiozaki, and M. Sato, *Topological Origin of Non-Hermitian Skin Effects*, Phys. Rev. Lett. **124**, 086801 (2020).

[30]  X. Zhang, T. Zhang, M.-H. Lu, and Y.-F. Chen, *A Review on Non-Hermitian Skin Effect*, Adv. Phys. X **7**, 2109431 (2022).

[31]  T. E. Lee, *Anomalous Edge State in a Non-Hermitian Lattice*, Phys. Rev. Lett. **116**, (2016).

[32]  S. Weidemann, M. Kremer, T. Helbig, T. Hofmann, A. Stegmaier, M. Greiter, R. Thomale, and A. Szameit, *Topological Funneling of Light*, Science **368**, 311 (2020).

[33]  L. Xiao, T. Deng, K. Wang, G. Zhu, Z. Wang, W. Yi, and P. Xue, *Non-Hermitian Bulk–Boundary Correspondence in Quantum Dynamics*, Nat. Phys. **16**, 761 (2020).

[34]  A. Ghatak, M. Brandenbourger, J. van Wezel, and C. Coulais, *Observation of Non-Hermitian Topology and Its Bulk–Edge Correspondence in an Active Mechanical Metamaterial*, Proc. Natl. Acad. Sci. **117**, 29561 (2020).

[35]  L. Zhang et al., *Acoustic Non-Hermitian Skin Effect from Twisted Winding Topology*, Nat. Commun. **12**, 6297 (2021).

[36]  D. Zou, T. Chen, W. He, J. Bao, C. H. Lee, H. Sun, and X. Zhang, *Observation of Hybrid Higher-Order Skin-Topological Effect in Non-Hermitian Topolectrical Circuits*, Nat. Commun. **12**, 7201 (2021).

[37]  W. Wang, X. Wang, and G. Ma, *Non-Hermitian Morphing of Topological Modes*, Nature **608**, 50 (2022).

[38]  Y. Yi and Z. Yang, *Non-Hermitian Skin Modes Induced by On-Site Dissipations and Chiral Tunneling Effect*, Phys. Rev. Lett. **125**, 186802 (2020).

[39]  K. Zhang, Z. Yang, and C. Fang, *Universal Non-Hermitian Skin Effect in Two and Higher Dimensions*, Nat. Commun. **13**, 2496 (2022).

[40]  T. Hofmann et al., *Reciprocal Skin Effect and Its Realization in a Topolectrical Circuit*, Phys. Rev. Res. **2**, 023265 (2020).

[41]  Z. Fang, M. Hu, L. Zhou, and K. Ding, *Geometry-Dependent Skin Effects in Reciprocal Photonic Crystals*, Nanophotonics **0**, (2022).

[42]  See the Supplemental Material for the properties of the GDSE, the experimental setup, and the retrieved parameters of the trapezoid and rectangular lattices, which includes Refs. [39].

[43]  W. Wang, X. Wang, and G. Ma, *Extended State in a Localized Continuum*, Phys. Rev. Lett. **129**, 264301 (2022).

[44]  K. Yokomizo and S. Murakami, *Non-Bloch Bands in Two-Dimensional Non-Hermitian Systems*, arXiv:2210.04412.

[45] H. Jiang and C. H. Lee, *Dimensional Transmutation from Non-Hermiticity*, arXiv:2207.08843.

[46] Dan S. Borgnia, Alex Jura Kruchkov, and Robert-Jan Slager, *Non-Hermitian Boundary Modes and Topology*, Phys. Rev. Lett. **124**, 056802 (2020).

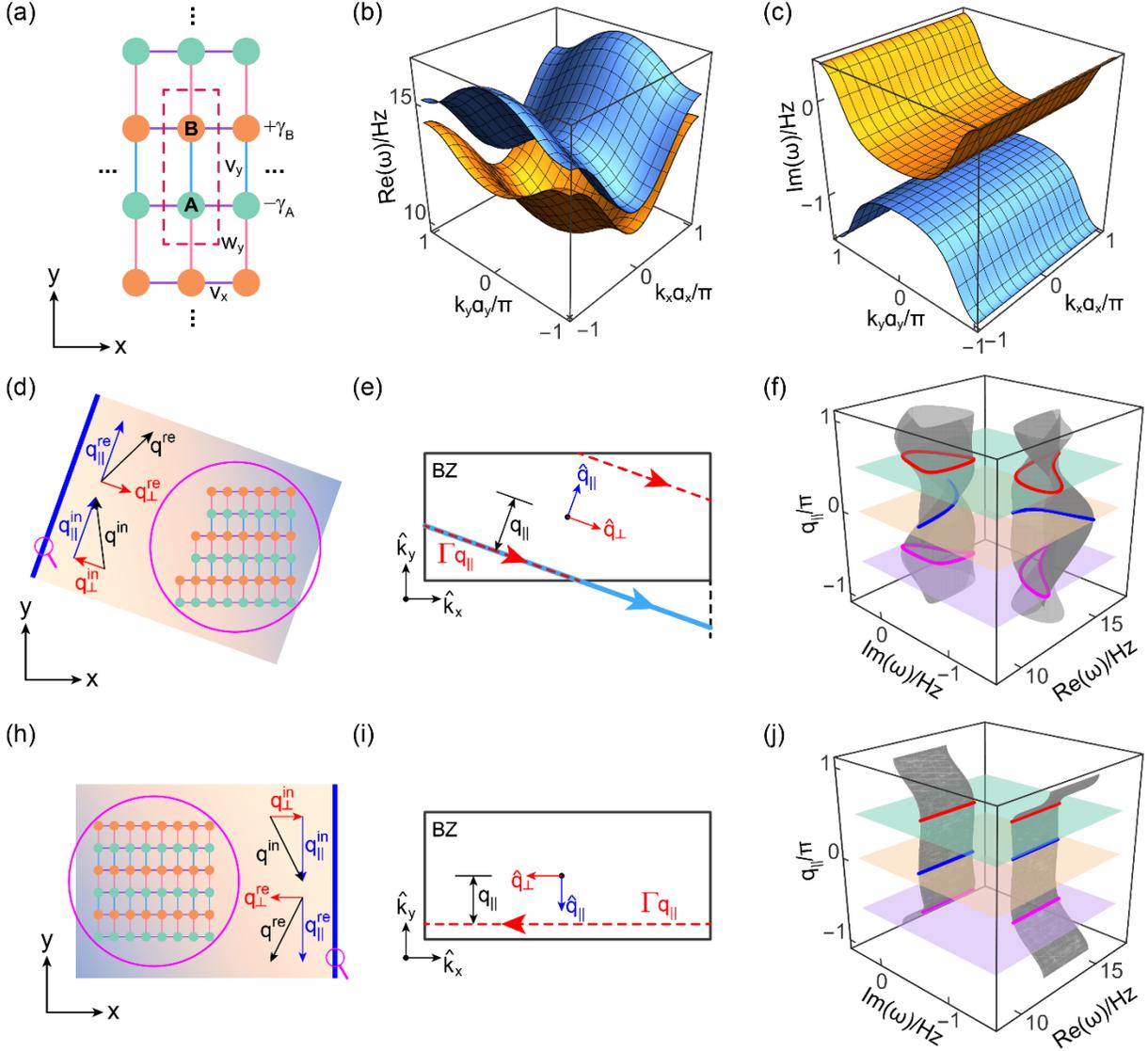

FIG.1 (a) The 2D non-Hermitian lattice. The dashed red box marks the unit cell. The blue, pink, and purple lines are the hopping terms $v_y$, $w_y$, and $v_x$, respectively. The real (b) and imaginary (c) parts of the PBC spectrum. The schematic of the reflection at (d) an oblique edge and (h) a vertical boundary. The magenta circles show the zoomed-in views of the lattice arrangement at the boundaries. $\Gamma_{q_\parallel}$ with a fixed $q_\parallel$ across the BZ for the oblique edge (e) and the boundary parallel to the $y$-axis (i). (f, j) The gray surfaces are the complex spectrum as a function of $q_\parallel$ for the cases shown in (e, f), respectively. The green, orange, and pink surfaces correspond to $q_\parallel = 0.5\pi$, $q_\parallel = 0$, and $q_\parallel = -0.6\pi$, respectively.

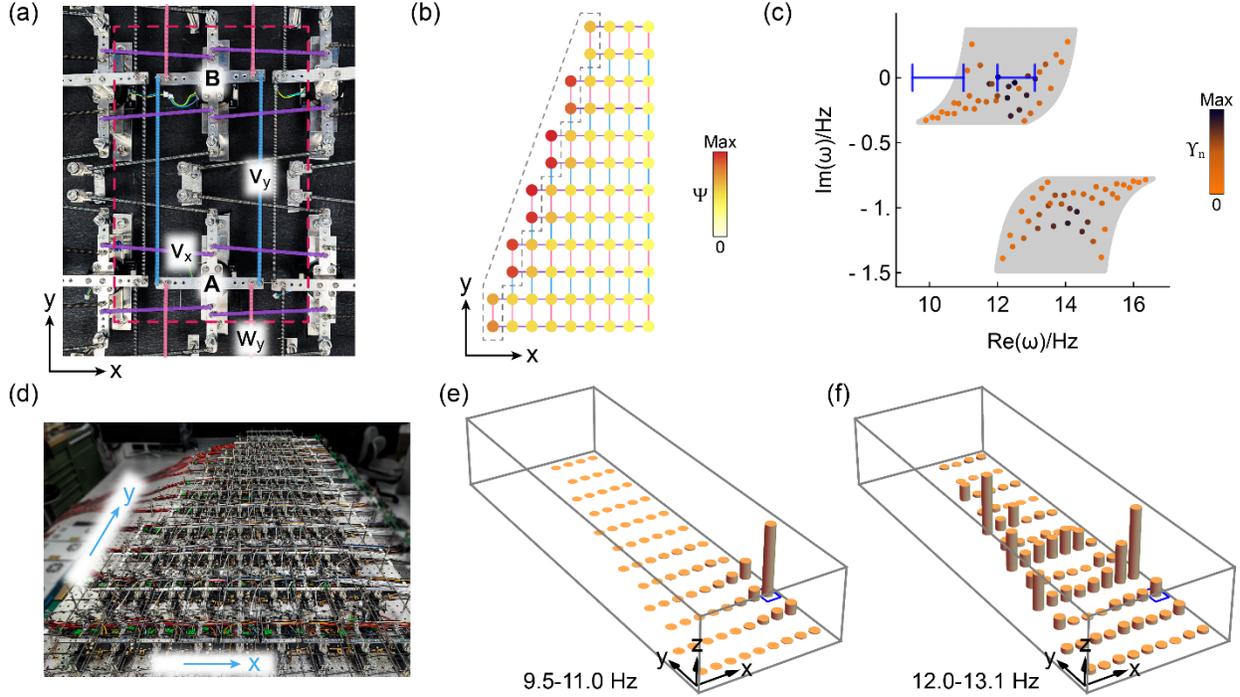

FIG. 2 (a) The photo of the coupled rotational oscillators realizing the unit cell shown in Fig. 1(a). (b) The schematic model of the trapezoid structure with 78 sites. The dashed line encloses the region $\mathcal{B}$. The color plot represents the spatial distribution of $\Psi(\mathbf{x})$. (c) The OBC complex spectrum (colorful dots) of the trapezoid structure. The gray background displays the corresponding PBC spectrum, and the color of the solid dots plot $\Upsilon_n$ for each eigenstate. (d) The photo of the coupled mechanical oscillator arranged in a trapezoid lattice. The plots of the spectral sum with different input chirp signals covering the frequency intervals (e) $\Delta f_1 = [12.0, 13.1]$ Hz and (f) $\Delta f_2 = [9.0, 11.0]$ Hz. The blue boxes in (e, f) mark the source.

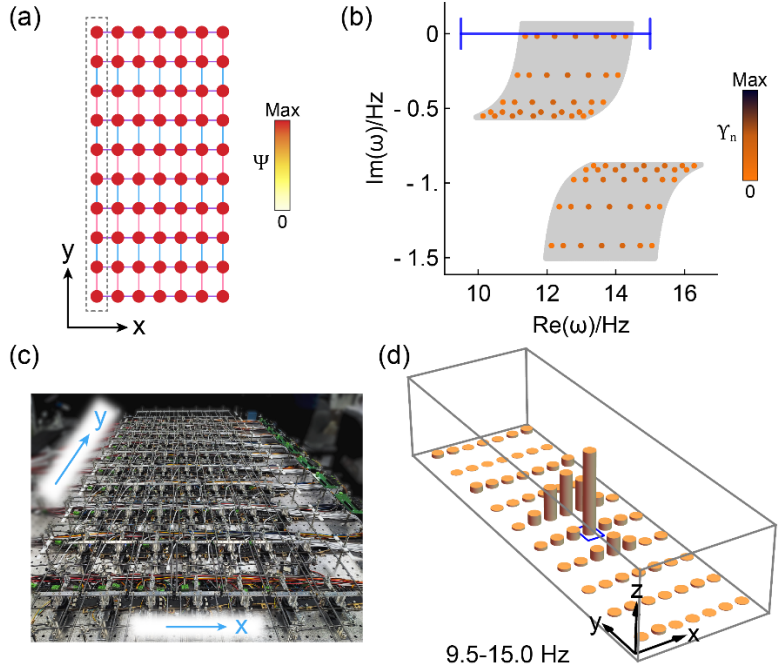

FIG. 3 (a) The schematic model of the rectangular structure with 70 sites. The gray dashed line encloses the region $\mathcal{B}$. The color plot represents the spatial distribution of $\Psi(\mathbf{x})$. (b) The OBC complex spectrum (colorful dots) of the rectangular structure. The gray background displays the corresponding PBC complex spectrum, and the color of the solid dots plot $\Upsilon_n$ for each eigenstate. The color bar is the same as that in Fig. 3(b). (c) The photo of the coupled mechanical oscillator arranged in a rectangular lattice. (d) The plots of the spectral sum with input chirp signals covering 9.5-15.0 Hz. The blue box marks the source.